\documentclass[letterpaper, 10 pt, conference]{ieeeconf}
\IEEEoverridecommandlockouts
\usepackage{amsmath}
\usepackage{amssymb}
\usepackage{siunitx}
\usepackage{dsfont}
\usepackage[ruled]{algorithm2e}
\DontPrintSemicolon

\usepackage{mathtools}
\mathtoolsset{showonlyrefs}

\newcommand{\R}{\mathbb{R}}

\newcommand{\Ex}{\mathbb{E}}
\newcommand{\ind}{1}
\newcommand{\initialset}{\ensuremath{\mathcal{X}_0} }

\newcommand{\distset}{\ensuremath{\mathcal{D}} }
\newcommand{\rs}{\ensuremath{R_{[t_0,t_1]}}}
\newcommand{\ars}{\ensuremath{\hat{R}_{[t_0,t_1]}}}

\usepackage{mathtools}
\DeclarePairedDelimiter{\ceil}{\lceil}{\rceil}

\newtheorem{remark}{Remark}
\newtheorem{problem}{Problem}
\newtheorem{lemma}{Lemma}

\title{\LARGE \bf
Data-Driven Reachability Analysis with Christoffel Functions
}

\author{ \parbox{5 in}{\centering Alex Devonport, Forest Yang, Laurent El
        Ghaoui, Murat Arcak\\
         Electrical Engineering and Computer Sciences\\
         University of California, Berkeley\\
     {\centering \tt\small \{alex\_devonport,forestyang,elghaoui,arcak\}@berkeley.edu}}
} 

\newtheorem{theorem}{Theorem}

\begin{document}

\maketitle
\thispagestyle{empty}
\pagestyle{empty}

\begin{abstract}
We present an algorithm for data-driven reachability analysis that estimates
finite-horizon forward reachable sets for general nonlinear systems using
level sets of a certain class of polynomials known as Christoffel
functions. The level sets of Christoffel functions are known empirically to
provide good approximations to the support of probability distributions: the
algorithm uses this property for reachability analysis by solving a
probabilistic relaxation of the reachable set computation problem. We also
provide a guarantee that the output of the algorithm is an accurate
reachable set approximation in a probabilistic sense, provided that a certain
sample size is attained. We also investigate three numerical examples to
demonstrate the algorithm's capabilities, such as providing non-convex
reachable set approximations and detecting holes in the reachable set.
\end{abstract}

\maketitle

\section{Introduction}

A popular and effective way to guarantee the safety of a system
in the face of uncertainty is \emph{reachability analysis}, a set-based method
that characterizes all possible evolutions of the system by computing reachable sets.
Many algorithms in reachability analysis use detailed
system information to compute a sound approximation to the reachable set, that
is an approximation guaranteed to completely contain (or be contained in) the
reachable set. However,
in many important applications, such as 
complex
cyber-physical systems that
are only accessible through simulations or experiments, this detailed
system information is not available, so these algorithms cannot be applied.

Applications such as these motivate \emph{data-driven} reachability analysis,
which studies algorithms to estimate reachable sets using the type of data
that can be obtained from experiments and simulations. 
These algorithms have the advantage of being able to estimate %
the
reachable sets of any system whose behavior can be simulated or measured
experimentally, without requiring any additional mathematical information about
the system. The main disadvantage of data-driven reachability algorithms is that
generally 
they
cannot provide the same type of soundness guarantees as
traditional reachability analysis algorithms; however, they can still 
guarantee accuracy of the estimates
in a
probabilistic sense with high confidence.

Data-driven reachability is a rapidly growing area of research within
reachability analysis. Many recent developments focus either on providing probabilistic
guarantees of correctness for data-driven methods that estimate the reachable
set directly from data,
for instance using results from statistical learning theory~\cite{devonport2020data} or scenario
optimization~\cite{marseglia2014hybrid,yang2016multi,ioli2017smart,sartipizadeh2019voronoi,hewing2019scenario,devonport2020estimating}. 
Others incorporate data-driven elements into more traditional reachability
approaches, for instance estimating entities such as discrepancy
functions~\cite{fan2017dryvr} or
differential inclusions~\cite{djeumou2020fly}. Finally, other developments include incorporating
data-driven reachability into verification tools for
cyber-physical systems~\cite{fan2017dryvr, qi2018dryvr}.

This paper investigates a data-driven reachability algorithm that directly
estimates the reachable set from data using the sublevel sets of an empirical
inverse Christoffel function, and provides a probabilistic guarantee of accuracy
for the method using statistical learning-theoretic methods. Christoffel
functions are a class of polynomials defined with respect to measures on $\R^n$:
a single measure defines a family of Christoffel function polynomials.
When the measure in question is defined by a probability distribution on $\R^n$
the level sets of Christoffel functions are known empirically to
provide 
tight
approximations to the support.
This support-approximating quality has motivated the use of Christoffel functions in several statistical applications, such as density estimation~\cite{lasserre2017empirical, lasserre2019empirical} and outlier detection~\cite{askari2018kernel}.
Additionally, the level sets have been shown, using the plug-in approach~\cite{cuevas1997plug}, to converge exactly to the support of the distribution (in the sense of
Hausdorff measure) when the degree of the polynomial approaches infinity, and
when the true probability distribution is available~\cite{lasserre2019empirical}.
When the true probability distribution is \emph{not} known, as is typically the case in
data analysis, the Christoffel function can be empirically estimated using a
point cloud of independent and identically distributed (iid) samples from the
distribution: this \emph{empirical Christoffel function} still provides accurate
estimates for the support, and some convergence results in this case are also known~\cite{pauwels2020data}.

The contribution of this paper is twofold. First, we provide an algorithm which
uses the level sets of a Christoffel function to estimate a reachable set using
a point cloud of iid samples from the reachable set, which can be obtained
through simulations by a 
Monte Carlo sampling scheme. 
Second, we provide a guarantee of the probabilistic accuracy of the reachable
set estimate produced by the algorithm: provided that a certain (finite) sample
size is attained, the level set provided by the algorithm is guaranteed to
achieve a user-specified level of probabilistic accuracy with high confidence.
Unlike the convergence results of~\cite{lasserre2019empirical,pauwels2020data}, this result holds for finite sample sizes and finite degrees.

\subsection*{Notation}

Given vectors $a,b\in\R^n$, a multidimensional interval (``interval'' for
brevity) is the set $[a,b]=\{x\in\R^n | a \le x \le b\}$, where $\le$ is the
standard partial order $\R^n$.
Given a vector $x$, a subscript $x_i$ denotes the $i^{th}$ element of $x$. Given
an ordered multiset of vectors (a collection of points in $\R^n$ for instance), a superscript $x^{(i)}$
denotes the $i^{th}$ member of the multiset.
For $x\in\R^n$, the vector $z_k(x)\in\R^{\binom{n+k}{n}}$ denotes the vector of monomials of degree $\le k$, including degree zero, evaluated at $x$: for instance, if $n=2$ and $k=2$, then $z_k(x)=[1~~x_1~~x_2~~x_1x_2~~x_1^2~~x_2^2]^\top$.
The space of polynomials of degree $\le d$ in $n$ variables is denoted $\R[x]^n_d$: note that elements of $z_d$, treated as polynomials, form a basis for $\R[x]^n_d$.

\section{Preliminaries}

\subsection{Probabilistic Reachability Analysis}
\label{subsec:probabilistic_reachability_analysis}

Consider a 
dynamical system with a state transition function
$\Phi(t_1;t_0, x_0, d)$
that maps an initial state $x(t_0)=x_0\in\R^n$ at time $t_0$ to a unique final state
at time $t_1$, under 
a disturbance $d:[t_0,t_1]\to\R^{w}$. 
For instance, when the system state dynamics $ \dot{x}(t) = f(t,x(t),d(t))$
are known and have unique solutions on the interval $[t_0,t_1]$, then
$\Phi(t_1;t_0, x_0, d)$ is just $x(t_1)$, where $x$ is the solution of the state dynamics with initial condition $x(t_0)=x_0$.
In addition to representing exogenous disturbances, the disturbance signal $d$
may account for deviations of an input from a nominal control law.

For the problem of forward reachability analysis, we are also given an
\emph{initial set} $\initialset\subset\R^n$, a set $\distset$ of allowed
disturbances and a time range $[t_0,t_1]$. The \emph{forward reachable set} is
then defined as the set of all states to which the system can transition in the
time range $[t_0,t_1]$ with initial states in $\initialset$ and disturbances in
$\distset$, that is the set
\begin{equation}
     \rs = \{\Phi(t_1;t_0,x_0, d) : x_0\in\initialset, d\in\distset\}.
\end{equation}

To tackle the problem of estimating the forward reachable set by statistical means, we
add probabilistic structure to the reachability problem that corresponds
to taking random independent samples from the reachable set. Specifically,
we take random variables $X_0$ and $D$ that take values on $\initialset$ and $\distset$ respectively. These random variables then
induce a random variable 
$\Phi(t_1;t_0, X_0, D)$ 
over the forward reachable
set, whose probability measure we denote as $\mu$.

\begin{remark}
    The random variables $X_0$ and $D$ may
have a physical significance, if the initial states, inputs, or disturbances are
known to behave randomly in the problem at hand. However, they do not need to:
they may be considered as \emph{instrumental distributions} whose purpose is to
provide a consistent rule for selecting initial states and disturbances
at random. 
\end{remark}

The measure $\mu(A)$ of a set $A\in\R^n$ has an intuitive interpretation: if we take samples $x_0$ and $d$ of the random
variables $X_0$ and $D$, then the vector $\Phi(t_1;t_0,x_0,d)$ lies in
$A$ with probability $\mu(A)$.
Additionally, the smallest set of measure 1 is the reachable set. 
This interpretation motivates $\mu(A)$ as a
measure of \emph{probabilistic accuracy}: if a set $A\subseteq\R^n$ has a
greater measure $\mu(A)$ than a set $B\subseteq\R^n$, then $A$ is a more
accurate approximation of the reachable set than $B$, in the sense that it
``misses'' less of the probability mass than $B$ does.
In the probabilistic
version of the forward reachability problem, our goal is to find
reachable set approximations $\ars$ such that $\mu(\ars)$ is close to 1.
Formally, we look to solve the following problem.
\begin{problem}\label{prb:frs_prob}
    Given the state transition function 
     $\Phi(t_1;t_0, x_0, u)$, time range $[t_0,t_1]$,
     initial set $\initialset$, and disturbance set
     $\distset$, the random variables $X_0$ and $D$, and an \emph{accuracy level}
    $\epsilon\in(0,1)$, compute a set $\ars$ such that $\mu(\ars) \ge 1-\epsilon$.
\end{problem}

Selecting a set with high measure under $\mu$ is not sufficient to
ensure a reasonable estimate, since the trivial solution $\ars=\R^n$ satisfies
$\mu(\ars)=1$. To avoid this problem we require some regularization, such as
requiring that $\ars$ be compact and penalizing estimates with high volume.

\subsection{Christoffel Functions}

Given a finite measure $\mu$ on $\R^n$ and a positive integer $k$,
the Christoffel function of order $k$ is defined as the ratio
\begin{equation}
    \label{eq:cfun_def}
    \kappa(x) = \frac{1}{z_k(x)^\top M^{-1}z_k(x)},
\end{equation}
where $M$ is the matrix of moments
\begin{equation*}
    M = \int_{\R^n}z_k(x) z_k(x)^\top d\mu(x)
\end{equation*}
and $z_k(x)$ is the vector of monomials of degree $\le k$.
We assume throughout that $M$ is positive definite, ensuring that $M^{-1}$
exists. The Christoffel function has several important application in
approximation theory, where its asymptotic properties are used to prove the
regularity and consistency of Fourier series of orthogonal polynomials. For our
purposes, it is more convenient to use the \emph{inverse Christoffel function}
\begin{equation}
    {\kappa(x)}^{-1} = z_k(x)^\top M^{-1}z_k(x),
\end{equation}
which is a polynomial of degree $2k$.
In Problem~\ref{prb:frs_prob}, and more generally in the problem of estimating
a probability distribution from samples, $\mu$ is a probability measure which we do not \emph{a priori} know. In this case, we instead use an empirical
estimate of $\mu$ constructed from a collection of independently and identically distributed (iid) samples $x^{(i)}$, $i=1,\dotsc,N$ samples from $\mu$, namely
\begin{equation*}
    \hat{\mu}=\frac{1}{N}\sum_{i=1}^N \delta_{x^{(i)}},
\end{equation*}
where $\delta_x$ is the \emph{Dirac measure} satisfying 
$\int f(y)d\delta_x(y)=f(x)$. The measure $\hat{\mu}$ itself defines a Christoffel function, whose inverse
\begin{equation}
    \label{eq:empirical_cfun_def}
    \begin{split}
    C(x)&=
    \hat{\kappa}^{-1}(x)=
    z_k(x)^\top \hat{M}^{-1}z_k(x)\\
    &= 
    z_k(x)^\top \left(
    \frac{1}{N}
    \sum_{i=1}^N 
    z_k(x^{(i)})
    z_k(x^{(i)})^\top
    \right)^{-1}z_k(x),
    \end{split}
\end{equation}
is called the \emph{empirical inverse Christoffel function}.
The matrix $\hat{M}$ is positive definite (and hence $\hat{M}^{-1}$ exists) if $N \ge \binom{n+k}{n}$ and the $x^{(i)}$ do not all belong to the zero set of a single degree $k$ polynomial.

\section{Christoffel Function Level Sets as Reachable Set Approximations}
\label{sec:christoffel_function_level_sets_as_reachable_set_approximations}

The ability of level sets of Christoffel functions to estimate the support of probability distributions motivates Algorithm~\ref{alg:cfun} as a data-driven strategy for solving Problem~\ref{prb:frs_prob}.
Specifically, Algorithm~\ref{alg:cfun} computes an empirical inverse Christoffel function $C(x)$ and a level parameter $\alpha\in\R$, and returns the sublevel set $\{x\in\R^n : C(x) \le \alpha\}$ as a proposed solution to Problem~\ref{prb:frs_prob}.

\begin{algorithm}
    \SetAlgoLined
    \caption{Data-driven reachable set estimation by a sublevel set of an
    empirical inverse Christoffel function.}
    \label{alg:cfun}
    \KwIn{Transition function $\Phi$ of a system with state dimension $n$;
random variables $X_0$ and $D$ defined on $\initialset$ and
$\distset$
respectively; time range $[t_0, t_1]$; probabilistic guarantee
parameters $\epsilon$ and $\delta$; Christoffel function order $k$.}
\KwOut{Set $\ars$ representing an $\epsilon$-accurate
reachable set estimate with confidence $1-\delta$.}
    Set number of samples
    \begin{equation*}
        N = \ceil*{
            \frac{5}{\epsilon}\left(
            \log\frac{4}{\delta} + \binom{n+2k}{n} \log\frac{40}{\epsilon}
    \right)}
    .
    \end{equation*}\;
    \ForAll{$i \in \{1,\dotsc,N\}$} {
        Take iid samples $x_0^{(i)}$ and $d^{(i)}$ from $X_0$ and $D$ respectively;\;
        evaluate $x_f^{(i)}=\Phi(t_1; t_0, x_0^{(i)}, d^{(i)})$.\;
    }
    Compute the matrix $\hat{M}^{-1}$ and level parameter $\alpha$, where
    \begin{equation*}
        \begin{aligned}
        \hat{M} &= \frac{1}{N} \sum_{i=1}^N z_k(x_f^{(i)})z_k(x_f^{(i)})^\top, \\
            \alpha &= \max_{i=1,\dotsc,N} z_k(x_f^{(i)})^\top \hat{M}^{-1} z_k(x_f^{(i)}).
        \end{aligned}
    \end{equation*}
    \;
    Record the set
    \begin{equation*}
        \ars=\{x\in\R^n : z_k(x)^\top \hat{M}^{-1} z_k(x) \le \alpha\}
    \end{equation*}
    as the reachable set estimate.\;
\end{algorithm}

Since Algorithm~\ref{alg:cfun} is a randomized algorithm, it is possible that a particular run will produce an invalid solution to Problem~\ref{prb:frs_prob}. However, Theorem~\ref{thm:cfun_pac} guarantees that the probability that this occurs is no greater than $\delta$, a parameter that the user can specify in advance.

\begin{theorem}
    \label{thm:cfun_pac}
    Let $C$ denote the empirical inverse Christoffel function for a point
    cloud $x^{(1)},\dotsc,x^{(N)}$ of iid samples from $\mu$, i.e.
    \begin{equation*}
        C(x)=z_k(x)^\top\left(\frac{1}{N}\sum_{i=1}^N z_k(x^{(i)})z_k(x^{(i)})^\top\right)^{-1}z_k(x),
    \end{equation*}
    and let $\alpha = \max_i C(x^{(i)})$.
    Let $\mu^N$ denote the joint probability measure corresponding to $N$ iid samples from $\mu$.
    If
    \begin{equation}
        \label{eq:cfun_pac_bound}
        N \ge 
            \frac{5}{\epsilon}\left(
            \log\frac{4}{\delta} + \binom{n+2k}{n} \log\frac{40}{\epsilon}
        \right),
    \end{equation}
    then
    \begin{equation}
    \begin{aligned}
    \mu^N\bigg(&\{(x^{(1)},\dotsc,x^{(N)}):\\ 
    &\mu\left(\{x\in\R^n : C(x) \le \alpha\}\right) \ge 1-\epsilon\}\bigg) \ge 1-\delta.
    \end{aligned}
    \end{equation}
    This means that, with probability $\ge 1-\delta$, the
    $\alpha$-sublevel set of $C(x)$ contains at least $1-\epsilon$ of the
    probability mass of $\mu$.
\end{theorem}
The probability $1-\delta$ is the \emph{confidence} that the solution is valid. For instance, suppose we set $\delta=10^{-9}$: then Theorem~\ref{thm:cfun_pac} gives us the confidence that there is less than a one in a billion chance that Algorithm~\ref{alg:cfun} will fail to solve Problem~\ref{prb:frs_prob}.

The proof of this result is based on the following two results from statistical
learning theory. 

\begin{lemma}[\cite{dudley1978central}, Theorem 7.2]
    \label{lem:pos_vc}
    Let $V$ be a vector space of functions $g:\R^n\to\R$ with dimension $m$.
    Then the class of sets
    \begin{equation*}
        \text{Pos}(V) = \left\{\ \{x | g(x)\ge 0\}, g\in V\right\}
    \end{equation*}
    has Vapnik–Chervonenkis (VC) dimension $m$.
\end{lemma}

\begin{lemma}[\cite{alamo2009randomized}, Corollary 4]
    \label{lem:generic_pac}
    Let $\mathcal{C}$ be a class of sets with VC dimension $m$. For a set
    $c\in\mathcal{C}$, let $\hat{\ell}(c)=\frac{1}{N}\sum_{i=1}^N
    \ind\{x^{(i)}\notin c\} $ be the empirical error from a sample of $M$ iid
    samples from $\mu$, and let $\ell(c)=\Ex_\mu[\ind\{X\notin c\}]=1-\mu(c)$ be the
    generalization error. If
    \begin{equation}
        \label{eq:generic_pac_bound}
        N \ge \frac{5}{\epsilon}\left(
            \log\frac{4}{\delta} + m \log\frac{40}{\epsilon}
        \right),
    \end{equation}
    and if $\hat{\ell}(c)=0$, that is if all of the points $x^{(i)},\ i=1,\dotsc,n$ are contained in the concept $c$, then $\mu^N\left(\{x^{(1)},\dotsc,x^{(N)} : \ell(c) \le \epsilon\}\right) \ge 1-\delta$.
\end{lemma}
Theorem~\ref{thm:cfun_pac} follows from Lemmas~\ref{lem:pos_vc} and~\ref{lem:generic_pac}
because the
set $c=\{x\in\R^n | C(x) \le \alpha\}$
belongs to the class $\mathcal{C}=\text{Pos}(\R[x]^n_d)$ and satisfies
$\hat{\ell}(c)=0$, and because the dimension of $\R[x]^n_d$ is $\binom{n+2k}{n}$.

In addition to providing a high-confidence solution to
Problem~\ref{prb:frs_prob}, Algorithm~\ref{alg:cfun} also achieves the
regularization goals mentioned at the end of Section~\ref{subsec:probabilistic_reachability_analysis}. In particular,
the estimate $\ars$ produced by Algorithm~\ref{alg:cfun} is compact, since it is
a sublevel set of the sum-of-squares polynomial $z(x)^\top \hat{M}^{-1} z(x)$.
Furthermore, the
level parameter $\alpha$ can equivalently be defined as the solution to the
optimization problem
\begin{equation*}
    \begin{aligned}
&\text{arg }\underset{\alpha > 0}{\text{min}}
& & \alpha & \\
& \text{subject to}
& & z_k(x^{(i)})^\top M^{-1} z_k(x^{(i)}) \le \alpha,\ i=1,\dotsc,N.
    \end{aligned}
\end{equation*}
In this problem, $\alpha$ acts as a penalty term for the volume of the sublevel
set, since the volume increases monotonically with increasing $\alpha$.

\begin{remark}
    \label{rmk:rs_subset}
    In some reachability problems, we are only interested in computing a
    reachable set for a subset of the state variables.
    For example, suppose the state is $(x_1,\dotsc,x_n)\in\R^n$, and we wish to
    verify a safety specification involving only the states $x_1,\dotsc,x_m$,
    where $m<n$: a reachable set for the states $x_1,\dotsc,x_m$ would suffice
    for this problem.
    In cases like this, Algorithm~\ref{alg:cfun} can be modified to use only the first
    $m$ elements of the samples $x_f^{(i)}$. The output of the algorithm is then
    an empirical inverse Christoffel function with domain $\R^m$ whose sublevel
    set $\ars$ estimates the reachable set for the reduced set of states.
    In the sequel, we refer to this application of Algorithm~\ref{alg:cfun} as the \emph{reduced-state variant} of Algorithm~\ref{alg:cfun}.
\end{remark}

\section{Examples}
\label{sec:numerical_examples}

This section demonstrates Algorithm~\ref{alg:cfun}'s ability to make
accurate estimates of forward reachable sets with three numerical examples. 
We demonstrate how the parallel nature of the algorithm can be leveraged to improve computation times by running all experiments on two computing platforms: (i) a laptop with 4 2.6 GHz cores; and (ii) an instance of the AWS EC2 computing platform \texttt{c5.24xlarge}, a virtual machine with 96 3.6 GHz cores. 

\subsection{Chaotic Nonlinear Oscillator}
\label{subsec:chaotic_nonlinear_oscillator}

The first example is a reachable set estimation problem for the nonlinear,
time-varying
system with dynamics
\begin{equation}
    \label{eq:duffing_dynamics}
    \begin{split}
        \dot{x} &= y \\
        \dot{y} &= -\alpha y + x - x^3 + \gamma\cos(\omega t),
    \end{split}
\end{equation}
with states $x,y\in\R$ and parameters $\alpha, \gamma, \omega\in\R$. This system
is known as the \emph{Duffing oscillator}, a nonlinear oscillator which exhibits
chaotic behavior for certain values of $\alpha$, $\gamma$, and $\omega$, for instance
\begin{equation}
    \begin{aligned}
        \alpha &= 0.05, & \gamma &= 0.4, & \omega &= 1.3.
    \end{aligned}
\end{equation}
The initial is the interval such that $x(0)\in[0.95, 1.05]$,
$y(0)\in[-0.05,0.05]$, and we take $X_0$ to be the uniform random variable over this interval. The time range is $[t_0,t_1]=[0,100]$.

We use Algorithm~\ref{alg:cfun}
to compute a reachable set for~\eqref{eq:duffing_dynamics} using an order $k=10$
empirical inverse Christoffel function with accuracy and confidence parameters
$\epsilon=0.05$, $\delta=10^{-9}$. 
With these parameters,~\eqref{eq:cfun_pac_bound} states that $N=156,626$ samples
are required to ensure that Theorem~\ref{thm:cfun_pac} holds for the reachable
set estimate. Total computation times for this example were 39 minutes on the laptop, and 41 seconds on \texttt{c5.24xlarge}.

\begin{figure*}
    \centering
    \includegraphics[width=0.8\linewidth]{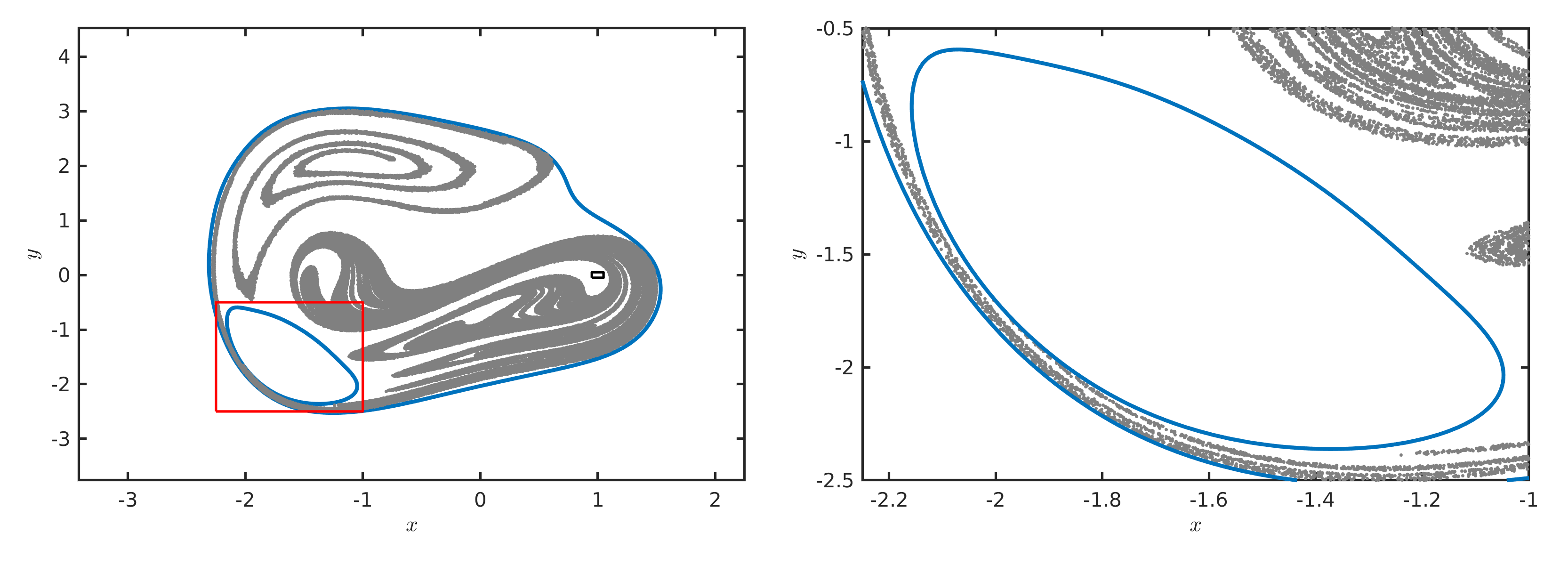}
    \caption{\emph{Left}: reachable set estimate for the Duffing oscillator
    system (blue contour), the cloud of 156,626 samples used to compute the
empirical inverse Christoffel function (grey points), and the initial set
(black box). \emph{Right}: enlarged version of the region in the left plot
enclosed by the red box, showing the region excluded from the reachable set.}%
    \label{fig:duffing_rs_zoom}
\end{figure*}

Figure~\ref{fig:duffing_rs_zoom} shows the reachable set estimate
for the Duffing oscillator system with the problem data given above, and the
point cloud of $156,626$ samples used to compute the empirical inverse
Christoffel function and the level parameter $\alpha$. The reachable set
estimate is neither convex nor simply connected, closely following the
boundaries of the cloud of points and excluding an empty region within the cloud of
points.

To experimentally verify that the assertion of Proposition~\ref{thm:cfun_pac}
holds for the reachable set estimate, we compute an \emph{a posteriori} estimate
of the accuracy of the empirical inverse Christoffel function sublevel set. To
do this, we first compute a new set of sample points of size $N_{ap}$. Denoting by $N_{out}$ the
number of new samples that lie outside of the reachable set estimate, we can
compute the empirical accuracy of a reachable set approximation as
$1-N_{out}/N_{AP}$. We use $N_{AP}=46,\!052$ sample points to make the \emph{a posteriori}
estimate. This sample size ensures that a one-sided Chernoff bound holds, which
guarantees that empirical accuracy is within 1\% of the true with 99.99\%
confidence. The \emph{a posteriori} empirical accuracy computed with this sample
is $1-(2\times 10^{-5})$, ensuring that the true accuracy of the reachable set
estimate is at least $0.99-2\times10^{-5}$ with 99.99\% confidence. This is well
in excess of the $0.95$ accuracy guaranteed by Theorem~\ref{thm:cfun_pac}.

\subsection{Planar Quadrotor Model}

The next example is a reachable set estimation problem for horizontal position
and altitude in a nonlinear model of the planar dynamics of a quadrotor used as
an example  in~\cite{mitchell2019invariant,bouffard2012board}. 
The dynamics for this model are
\begin{equation}
    \begin{split}
        \ddot{x} &= u_1 K\sin(\theta)\\
        \ddot{h} &= -g + u_1 K\cos(\theta) \\
        \ddot{\theta} &= -d_0\theta - d_1\dot{\theta} + n_0 u_2, \\
    \end{split}
\end{equation}
where $x$ and $h$ denote the quadrotor's horizontal position and altitude in
meters,
respectively, and $\theta$ denotes its angular displacement (so that the
quadrotor is level with the ground at $\theta=0$) in radians. The system has 6
states, which we take to be $x$, $h$, $\theta$, and their first derivatives. The
two system inputs $u_1$ and $u_2$ (treated as disturbances for this example) represent the motor thrust and the desired angle,
respectively. The parameter values used (following~\cite{bouffard2012board}) are $g=9.81$,
$K=0.89/1.4$, $d_0=70$, $d_1=17$, and $n_0=55$. The set of initial states is the interval such that 
\begin{equation*}
    \begin{aligned}
        x(0)&\in[-1.7, 1.7], & \dot{x}(0)&\in[-0.8, 0.8], \\
        h(0)&\in[0.3, 2.0], & \dot{h}(0)&\in[-1.0, 1.0], \\
        \theta(0)&\in[-\pi/12, \pi/12], & \dot{\theta}(0)&\in[-\pi/2, \pi/2],
    \end{aligned}
\end{equation*}
the set of inputs is the set of constant functions $u_1(t)=u_1$, $u_2(t)=u_2$ $\forall t\in[t_0,t_1]$, whose values lie in the
interval
\begin{equation*}
    \begin{aligned}
        u_1&\in[-1.5+ g/K, 1.5 + g/K], & u_2&\in[-\pi/4, \pi/4],
    \end{aligned}
\end{equation*}
and we take $X_0$ and $D$ to be the uniform random variables defined over these intervals.
The time range is $[t_0,t_1]=[0,5]$. We take probabilistic parameters $\epsilon=0.05$, $\delta=10^{-9}$.
Since the goal of this example is to estimate a reachable set for the horizontal
position and altitude only, we are interested in a reachable set for a subset of
the state variables, namely $x$ and $h$.
As mentioned in Remark~\ref{rmk:rs_subset},
Algorithm~\ref{alg:cfun} can be used to estimate a reachable set for $x$ and $h$
in two ways: we can either compute a Christoffel function estimate for the
reachable set and take the ``shadow projection'' of the estimate onto $x$ and
$h$, or we could compute a Christoffel function estimate for $x$ and $h$
directly using the reduced-state variant of Algorithm~\ref{alg:cfun} with the $(x,h)$ components of
the reachable set data. To compare the relative accuracy and computational
expense of these methods, we compute a reachable set estimate for $(x,h)$ using
both methods.

\begin{figure}[htbp]
    \centering
    \includegraphics[width=0.9\linewidth]{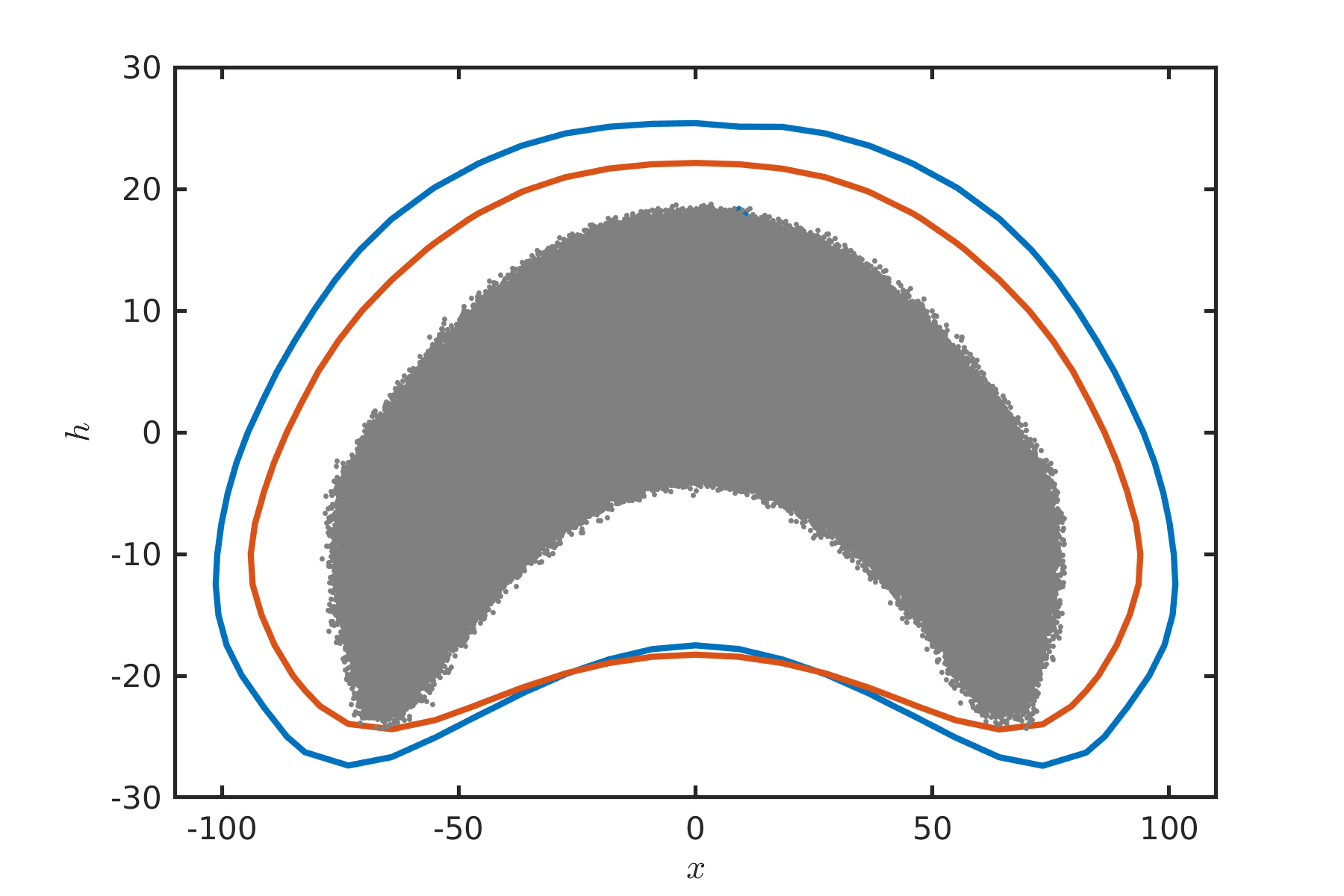}
    \caption{Reachable set estimates for the horizontal position and altitude of
    the planar quadrotor model, computed by projecting the output of
Algorithm~\ref{alg:cfun} onto $(x,h)$ (blue) and using the modification of
Algorithm~\ref{alg:cfun} mentioned in Remark~\ref{rmk:rs_subset}, where
the algorithm is run using only the $(x,h)$ components of the data (orange).}%
    \label{fig:quad_rs}
\end{figure}

Figure~\ref{fig:quad_rs} shows the reachable set estimates computed using both
methods using order $k=4$ inverse empirical Christoffel functions. Both
reachable estimates turn out to be similar, though the estimate using the
modification of Remark~\ref{rmk:rs_subset} is slightly tighter and significantly less
computationally expensive. Running Algorithm~\ref{alg:cfun} with the full state dimension $n=6$ and order $k=4$ with the $\epsilon$ and $\delta$ above requires $N=2,\!009,\!600$ samples: using the reduced-state variant brings the effective state dimension to $n=2$, and the sample size to $N=32,\!292$. The computation times in the full-state case were 77 minutes on the laptop and 2 minutes on \texttt{c5.24xlarge}; in the reduced-state case, computation times were 78 seconds on the laptop and 2 seconds on \texttt{c5.24xlarge}.
This shows that Algorithm~\ref{alg:cfun}'s ability to
work on subsets of the state space can speed up computations in cases where only
a subset of state variables are of interest.

\subsection{Monotone Traffic Model}

The final example is a special case of a continuous-time road traffic analysis problem
used as a reachability benchmark in~\cite{coogan2018benchmark,meyer2019tira,devonport2020pirk}. This problem
investigates the density of traffic on a single lane over a time range over four 
periods of duration $T$ using a 
discretization of the cell transmission model that divides the road into $n$
equal segments. The spatially discretized model
is an $n$-dimensional dynamical system with states $x_1,\dotsc,x_n$, where $x_i$ represents the
density of traffic in the $i^{th}$ segment.
Traffic enters segment through $x_1$ and flows through each successive segment before leaving through
segment $n$.
The state dynamics are
\begin{equation}
\begin{split}
    \label{eq:traffic_dynamics}
    \dot{x}_1 &= \frac{1}{T}\left(d-\min(c, vx_{1}, w(\overline{x}-x_{2}))\right)\\
    \dot{x}_i &= \frac{1}{T}\big( \min(c, vx_{i-1}, w(\overline{x}-x_{i})) \\
              &- \min(c, vx_{i}, w(\overline{x}-x_{i+1}))\big), \quad(i=2,\dotsc,n-1)\\
    \dot{x}_{n} &= \frac{1}{T}\left(\min(c, vx_{n-1}, w(\overline{x}-x_{n})/\beta) - \min(c, vx_{n}))\right),
\end{split}
\end{equation}
where $v$ represents the free-flow speed of
traffic, $c$ the maximum flow between neighboring
segments, $\bar{x}$ the maximum occupancy of a segment, and $w$ the congestion
wave speed. The input $u$ represents the influx of traffic into the first node.
For the reachable set estimation problem, we use a model with $n=6$ states, and take $T=30$, $v=0.5$, $w=1/6$, and $\bar{x}=320$. The initial set is the
interval such that $x_i(0)\in[100,200]$, $i=1,\dotsc,n$, the set of
disturbances is the set of constant disturbances with values in the range range $d\in[40/T, 60/T]$, and $X_0$ and $D$ are the uniform random variables over these sets. The time range is $[t_0, t_1]=[0, 4T]$.

The system dynamics~\eqref{eq:traffic_dynamics} are \emph{monotone}, or
order-preserving, meaning that if two initial conditions $x^{(1)}(0)$, $x^{(2)}(0)$ and
disturbances $d^{(1)}, d^{(2)}$ satisfy $x^{(1)}(0)\le x^{(2)}(0)$ (where $\le$ is the standard
partial order) and $d^{(1)}(t) \le d^{(2)}(t),\ t\in[0,T]$, then $x^{(1)}(T)\le x^{(2)}(T)$. 
This monotonicity allows for a convenient interval over-approximation of the
reachable set. If $\underline{x}$, $\overline{x}$ are the lower and upper bounds
of the interval of initial states, and $\underline{d}$, $\overline{d}$
are the lower and upper bounds on the values admitted by the disturbance signal, then
$[\Phi(t_1;t_0, \underline{x}, \underline{d}),
\Phi(t_1;t_0, \overline{x}, \overline{d})]$
is the smallest interval that contains the entire reachable set. While this
over-approximation is easy to compute, and the best possible over-approximation
by an interval, it is in general a conservative over-approximation because
reachable set may only occupy a small volume of the interval. Since the
empirical Inverse Christoffel function method can accurately detect the geometry
of the reachable set, we use this method to compare the shape of the reachable
set to the best interval over-approximation. In particular,
we use the reduced-state variant of Algorithm~\ref{alg:cfun}
to compute a reachable set for the traffic densities $x_5$ and $x_6$ at the
end of the road,
using an order $k=10$
empirical inverse Christoffel function with accuracy and confidence parameters
$\epsilon=0.05$, $\delta=10^{-9}$.
Computation times for this example were 10 minutes on the laptop and 2 minutes on \texttt{c5.24xlarge}.

\begin{figure}[htpb]
    \centering
    \includegraphics[width=0.9\linewidth]{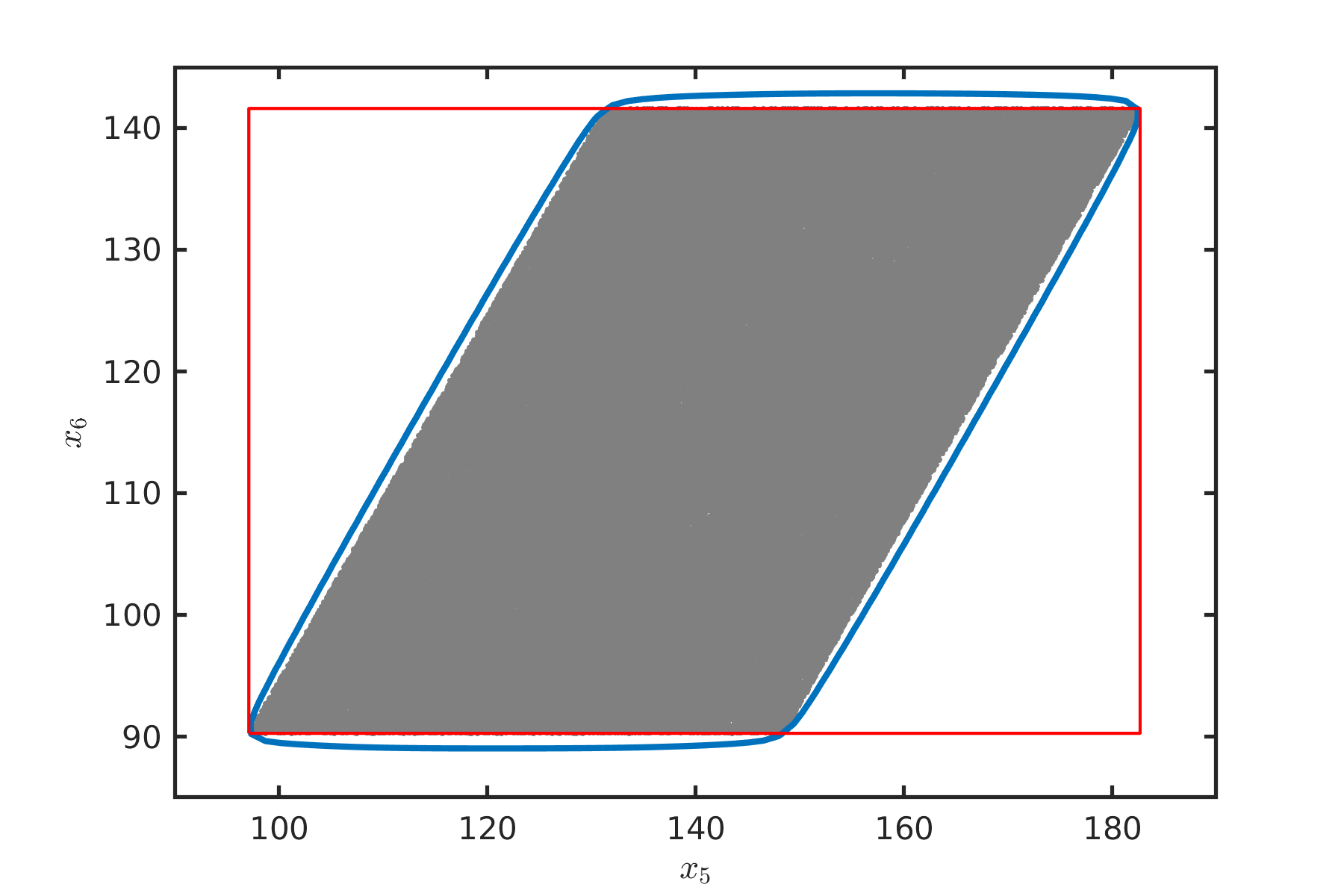}
    \caption{Reachable set estimate for the monotone traffic model with an order
    10 empirical inverse Christoffel function (blue), compared to the tight interval
       over-approximation (red). The reachable set estimate was computed with
       Algorithm~\ref{alg:cfun} using samples projected onto states $x_5$ and $x_6$.}%
       \label{fig:traf_rs}
\end{figure}

Figure~\ref{fig:traf_rs} compares the reachable set estimate computed with
Algorithm~\ref{alg:cfun} to the projection of the tight interval
over-approximation computed using the monotonicity property of the traffic
system.
The figure indicates that the tight interval over-approximation of the
reachable set is a somewhat conservative over-approximation, since the reachable
set has approximately the shape of a parallelotope whose sides are not axis-aligned.

\section{Conclusion}

Algorithm~\ref{alg:cfun} demonstrates that Christoffel functions, in addition to being useful in data analysis, can also be used as tools to provide principled, data-driven solutions to control-theoretic problems. While Theorem~\ref{thm:cfun_pac} assures that the proposed algorithm is a sound approach to solving reachability problems with data, and the examples of Section~\ref{sec:numerical_examples} demonstrate that the algorithm can provide accurate reachable set approximations, we believe it represents only the first step in applying Christoffel functions to data-driven reachability. For instance, the \emph{a posteriori} analysis of Section~\ref{subsec:chaotic_nonlinear_oscillator} suggests the sample bound of Theorem~\ref{thm:cfun_pac} is conservative, and could be significantly improved by applying some of the special properties of Christoffel functions. 

In addition, this paper did not explore how kernel methods can be used alongside Christoffel functions.
 Although we have defined the Christoffel function using the standard monomial basis vector $z_k(x)$,
the Christoffel function is in fact invariant to changes in polynomial
coordinates. For instance, $z_k(x)$ could be replaced with the feature vector
$\phi_k(x)$ of the polynomial kernel $(1+x^\top x)^k$, that is the monomial vector
$\phi_k(x)$ such that $\phi(x)^\top\phi(x)=(1+x^\top x)^k$. By an application of the kernel trick, this approach can be extended to kernels with infinite-dimensional feature spaces, as in~\cite{askari2018kernel}. However, the statistical learning-theoretic proof in this paper covers only the finite-dimensional case: providing finite-sample statistical guarantees for the infinite-dimensional case is a topic for future research.

\section*{Acknowledgments}

\begin{flushleft}
This work was supported in part by the grants ONR N00014-18-1-2209, AFOSR 
FA9550-18-1-0253, NSF ECCS-1906164.
\end{flushleft}

\bibliographystyle{IEEEtran}
\bibliography{IEEEabrv,refs}

\end{document}